\documentclass[aps,pra,twocolumn,showpacs,longbibliography,superscriptaddress]{revtex4-1}

\usepackage{mathtools,amssymb,graphicx,units,bm}
\usepackage{amsmath}
\usepackage[plainpages=false,pdfpagelabels,colorlinks=true,linkcolor=red,urlcolor=PineGreen,citecolor=PineGreen,pdftitle={Title},pdfauthor={},pdfdisplaydoctitle=true,pdfduplex=DuplexFlipLongEdge]{hyperref}
\usepackage{soul} 
\usepackage[dvipsnames,usenames]{color}
\usepackage[normalem]{ulem}
\usepackage{color}
\usepackage{bbold} 
\usepackage{float}
\emergencystretch=\maxdimen
\usepackage{dcolumn}
\usepackage{bm}
\usepackage{physics}%
\usepackage[dvipsnames]{xcolor}
\usepackage{bbold} 
\usepackage{float}
\emergencystretch=\maxdimen
\def\s{\sigma}
\def\ave#1{\langle #1\rangle}
\newcommand{\avs}{\ave{\text{sign}}}
\newcommand{\iv}{\mathbf{i}}
\newcommand{\jv}{\mathbf{j}}
\newcommand{\qv}{\mathbf{q}}
\newcommand{\rv}{\mathbf{r}}
\newcommand{\avd}{\left \langle D \right \rangle}
\newcommand{\up}{\uparrow}
\newcommand{\dn}{\downarrow}

\begin{document}

\normalem

\title{The extended Hubbard model on a honeycomb lattice}
\author{Welberth Kennedy}
\affiliation{Instituto de F\'\i sica, Universidade Federal do Rio de Janeiro
Cx.P. 68.528, 21941-972 Rio de Janeiro RJ, Brazil}
\author{Sebasti\~ao dos Anjos \surname{Sousa-J\'unior}}
\affiliation{Department of Physics, University of Houston, Houston, Texas 77004}
\author{Natanael C. Costa}
\author{Raimundo R. \surname{dos Santos}} 
\affiliation{Instituto de F\'\i sica, Universidade Federal do Rio de Janeiro
Cx.P. 68.528, 21941-972 Rio de Janeiro RJ, Brazil}

\begin{abstract}
The lack of both nesting and a van Hove singularity at half filling, together with the presence of Dirac cones makes the honeycomb lattice a special laboratory to explore strongly correlated phenomena. 
For instance, at zero temperature the repulsive [attractive] Hubbard model only undergoes a transition to an antiferromagnetic [$s$-wave superconducting degenerate with charge density wave (SC-CDW)] for sufficiently strong on-site coupling, $U/t\gtrsim 3.85$ [$U/t\lesssim -3.85$]; in between these, the system is a semi-metal, by virtue of the Dirac cones.
The addition of an additional  interaction, $V>0$ or $V<0$, between fermions in nearest neighbor orbitals should break the SC-CDW degeneracy giving rise to a phase diagram quite distinct from the one for the square lattice.
Here we perform determinant quantum Monte Carlo simulations to investigate the whole phase diagram,  covering the four combinations of signs of $U$ and $V$; the use of complex Hubbard-Stratonovich fields renders the region $|V|\leq |U|/3$ free from the `minus sign problem'. 
We calculate structure factors associated with different orderings, which, together with the double occupancy and the average sign allows us to map out the whole phase diagram. 
We have found that the SM phase forms a zone from which ordered phases are excluded, preventing the stabilization of a $d$-wave SC phase, i.e., only $s$-wave pairing is allowed.

\end{abstract}

\maketitle

\section{Introduction}
\label{sec:introduction}

The competition between interactions in itinerant fermionic systems usually gives rise to a multitude of interesting phases \cite{Fradkin2015}. For instance, while magnetic correlations arise from Coulomb repulsion between electrons, coupling with phonon degrees of freedom may induce charge and pairing correlations. 
From a theoretical perspective, the Extended Hubbard Model (EHM) is one of the simplest models describing emergent magnetic, charge, and superconducting orders in electronic systems.
Indeed, while a repulsive on-site interaction, $U>0$, favors the formation of magnetic moments and their effective coupling, the presence of a repulsive nearest-neighbor interaction, $V>0$, enhances charge correlations, thus favoring arrangements of doubly occupied sites.   
On a half-filled square lattice, an attractive $V<0$ gives rise to a $d$-wave superconducting state within a range of values of $U>0$ \cite{zhang89,Dagotto94,Yao22,chen2023,SousaJr2024}; upon doping, this model is expected to have strong bearings on high-$T_c$ superconductivity in the cuprates \cite{Scalapino92}.
This model has recently been realized in contexts such as the quasi-one-dimensional cuprate  Ba$_{2-x}$Sr$_x$CuO$_{3+\delta}$ \cite{Chen2021}, ultracold atoms in optical lattices \cite{Baier2016,Su23}, and in few-dopant quantum dot systems \cite{Wang2022}.

Phases emerging from competing interactions are often sensitive to the underlying lattice geometry.
A simple example is provided by the half-filled Hubbard model (HM) on a kagome lattice with anisotropic hopping: a crossover between kagome and Lieb lattices involves phases such as ferromagnetic insulating,  paramagnetic metallic, and Mott insulating~\cite{Lima2023,Yamada2011}. 
The recent discovery of correlated insulating phases and superconductivity in magic-angle twisted bilayer graphene (TBG)~\cite{Cao2018a,Cao2018b} has spawned renewed interest in the study of strongly correlated phenomena on a honeycomb (HC) lattice. 
Indeed, while electron-electron interactions had previously been invoked to explain the occurrence of a strain-driven semimetal-insulator transition in graphene \cite{SungHoon2012,ChenSi2016,HoKin2015}, a recent study pointed towards the potential control of electron-electron interaction in graphene through proximity screening \cite{kim2020}. 
In addition, some transition-metal dichalcogenides (TMD) with low-dimensional hexagonal-like lattice geometries have been found to exhibit charge-density wave and superconducting phases~\cite{Joe2014,Manzeli2017}. 
A Hubbard-like model for magic-angle TBG has been proposed \cite{kerelsky2019}, whose low-energy properties are strongly influenced by electron-electron (e-e) interactions, thus opening the possibility of novel many-body effects \cite{Vahedi2021}. 
Further, recent Matrix-RPA calculations suggest that the superconducting phase in the magic angle TBG is strongly influenced by  local and extended e-e interactions~\cite{braz2024}. 
Finally, we stress that  the properties of the HM on the HC lattice are quite different from those on the square lattice: the former displays a semimetallic-antiferromagnetic (SM-AFM) transition at a \emph{finite} value of $U$~\cite{Paiva2005,Assaad2013,Otsuka2016}; 
that is, the system goes from a gapless Dirac SM to an insulating magnetically ordered phase with a characteristic Mott gap. 

In view of the above, a thorough study of the EHM on a HC lattice is certainly of interest.
We note that the studies so far \cite{Wu2014,Faye2015,Schuler2018}
have been mean-field--based and focused on the first quadrant of the $U$-$V$ parameter space, namely $U,V>0$.
The picture that emerged at half filling is that in addition to the AFM and SM phases, already present in the HM, a finite $V>0$ induces the appearance of a charge-density wave (CDW) phase for $V\gtrsim U/3$; upon doping, superconducting phases may be stabilized \cite{Faye2015}.
One should have in mind that similar half-filling physics occurs in the Hubbard-Holstein model~\cite{Costa2021}, as a result of enhanced charge correlations   caused by the electron-phonon coupling. 
For completeness, we mention that several variants of the EHM on a HC lattice have also predicted the appearance of CDW phases  \cite{Kotov2012,Wu2014,golor2015,Terletska21},  as well as bond-ordered phases~\cite{Xu2018}, and non-trivial topological properties \cite{Grushin2013}. 

Our purpose here is to explore the full parameter space of the half-filled EHM on a HC lattice. 
Through determinant quantum Monte Carlo (DQMC) simulations, we will focus on the nature of the emerging phases, and on an accurate determination of the different phase boundaries. 
The layout of the paper is as follows. In Sec.\,\ref{sec:modelmethods}, we present the Hamiltonian and highlight the DQMC method. Section\,\ref{sec:results} presents the results for the ground-state transitions. Finally, Sec.\,\ref{sec:conclusions} summarizes our findings.

\section{Model and Methods}
\label{sec:modelmethods}

The Extended Hubbard model (EHM) Hamiltonian reads
\begin{align}
    \mathcal{H}   = & -t \sum_{\langle \textbf{i,j} \rangle} (c_{\textbf{i}\sigma}^\dagger c^{\phantom{\dagger}}_{\textbf{j}\sigma} + \text{H.c.})  - \mu \sum_{\textbf{i},\sigma} n_{\textbf{i},\sigma}\nonumber\\
    & + U\sum_\textbf{i} n_{\textbf{i}\uparrow}n_{\textbf{i}\downarrow} +V\sum_{\langle \textbf{i,j} \rangle} n_{\textbf{i}}n_{\textbf{j}}, 
    \label{eq:EHM}
\end{align}
with $c_{\iv,\s}^{\dagger} (c_{\iv,\s}^{\phantom{\dagger}})$ being the creation (annihilation) operator for an electron with spin $\s (= \uparrow , \downarrow)$ on site $\iv$. 
Here, the sums run over sites of a two-dimensional honeycomb lattice with periodic boundary conditions and $\ave{\iv,\jv}$ denotes nearest-neighbor sites only. 
The first term in Eq.\,\eqref{eq:EHM} describes fermionic hopping, the second controls the band filling through the chemical potential, $\mu$, with $n_{\iv,\s}\equiv c_{\iv,\s}^{\dagger}c_{\iv,\s}^{\phantom{\dagger}}$. 
The third and fourth terms describe on-site and extended interactions, with strengths $U$ and $V$, respectively. 
Hereafter, we set the chemical potential to $\mu=U/2+3V$ to yield a half-filled band, and the energies are expressed in units of $t$. 

We investigate the ground state properties of Eq.\,\eqref{eq:EHM} using determinant quantum Monte Carlo simulations\,\cite{Blankenbecler81,Hirsch83, hirsch85,Kawashima2002,rrds2003,sorella2017}. 
In this method, the non-commutation between the kinetic and interaction terms of the Hamiltonian are taken care of through  the Suzuki-Trotter decomposition, which introduces an imaginary time dimension, $L_{\tau} = \beta / \Delta \tau$, where $\beta$ is the inverse temperature, and $\Delta \tau$ the discrete time step. 
To express the (quartic) interaction terms in a quadratic form we employ three different Hubbard-Stratonovich (HS) transformations, which introduce auxiliary fields, as discussed in Refs.\,\onlinecite{zhang89,golor2015,Yao22}. 
The trace over fermionic degrees of freedom can then be carried out by importance sampling of the auxiliary fields, where a product of determinants becomes the weight of configurations. 
At this point, we note that for the half-filled HM the product of determinants is positive definite, so that the simulations are free from the infamous minus-sign problem \cite{hirsch85,scalettar89,Kawashima2002,rrds2003,sorella2017}.
However, simulations for any $V\neq0$ with real HS fields suffer from the minus-sign problem; i.e., averages become extremely noisy as the average sign of the product of fermionic determinants, $\avs$, decreases.
Nonetheless, by using complex HS fields \cite{golor2015,Yao22,SousaJr2024}, we may extend the sign-free region to $|V|\leq |U|/z$ for the EHM on a lattice with coordination number $z$, $z=3$ and 4 for the HC and square lattices, respectively.
We note that when $U>0$, this complex HS mitigation can only be implemented on particle-hole symmetric Hamiltonians and at half-filling; when $U<0$ particle-hole symmetry can be relaxed, since the fermionic determinants are real and equal.
Further, since our simulations are not restricted to this range, we calculate $\avs$, which will play an auxiliary role in our analyses, as presented below. 

We examine charge and  magnetic correlations respectively through the staggered charge structure factor, 
\begin{equation}
    S_{\text{cdw}}(\qv)\! =\! \frac{1}{N} \sum_{\rv_{\iv}, \rv_{\jv}} e^{-i\qv \cdot (\rv_{\iv} - \rv_{\jv})} \!\langle(n_{A,\rv_{\iv}}\!\! -\! n_{B,\rv_{\iv}})(n_{A,\rv_{\jv}}\!\! - \!n_{B,\rv_{\jv}}) \rangle,
\label{eq:scdw}
\end{equation}
and the antiferromagnetic structure factor,
\begin{equation}
    S_{\text{afm}}(\qv)\! = \!\frac{1}{N} \sum_{\rv_{\iv}, \rv_{\jv}} e^{-i\qv \cdot (\rv_{\iv} - \rv_{\jv})} \!\langle(S^{z}_{A,\rv_{\iv}}\!\! -\! S^{z}_{B,\rv_{\iv}})(S^{z}_{A,\rv_{\jv}}\!\! -\! S^{z}_{B,\rv_{\jv}}) \rangle,
\label{eq:ss}
\end{equation}
where $n_{\lambda,\rv_{\iv}} = n_{\lambda,\rv_{\iv}\up} + n_{\lambda,\rv_{\iv}\dn}$, and $S^{z}_{\lambda,\rv_{\iv}} = n_{\lambda,\rv_{\iv}\up} - n_{\lambda,\rv_{\iv}\dn}$. Here $\lambda =  A,B $ labels the sublattices, while $\rv_{\iv}$ is the unit cell position of a given site $\iv$ and $N=2(L\times L)$ is the total number of sites, with $L$ being the linear size of the lattice.
We also examine superconducting correlations through the pairing structure factor,
\begin{align}
S_{\text{sc}}^{\alpha} (\qv)= \frac{1}{N} \sum_{\textbf{i,j}} e^{-i\qv \cdot (\rv_\iv - \rv_\jv)} \langle (\Delta _{A,\iv}^{\alpha} + \Delta_{B,\iv}^{\alpha}) (\Delta _{A,\jv}^{\alpha\dagger} + \Delta_{B,\jv}^{\alpha\dagger}) \rangle,
\label{eq:pairing_cor}
\end{align}
with 
\begin{equation}
	\Delta _{\lambda,\iv}^{\alpha} = \sum_\textbf{a} f^{\alpha} (\textbf{a}) c_{\textbf{i}\downarrow} c_{\textbf{i+a}\uparrow}, 
\end{equation}
where $f^{\alpha} (\textbf{a})$ is the form factor for a given pair-wave symmetry, $\alpha={s,s^{*},d,d^{*},p,p^{*},f}$~\cite{Xu2016,Ying2018,white89}. 

Phase boundaries and associated critical exponents may also be determined through the correlation ratio  \cite{Kaul2015,Sato2018,Liu2018,Darmawan2018,Otsuka2020},
\begin{align}
    R_{\gamma}(L,V) = 1- \frac{S_{\gamma}(\qv + \delta \qv)}{S_{\gamma}(\qv)},
    \label{eq:RgammaLV}
\end{align}
where $\qv = (0,0)$, $ \delta \qv$ is chosen as its nearest- or next-nearest--neighbor reciprocal wave vector, and $\gamma$ denotes ``cdw'', ``afm'' or ``sc''. 
On the HC reciprocal lattice, the two nonequivalent nearest neighbors of $\qv = (0,0)$ are $\delta^{(0,1)}_{\qv} = \widehat{\mathbf{b}}_{2}/L$ and $\delta^{(1,1)}_{\qv} = (\widehat{\mathbf{b}}_{1} + \widehat{\mathbf{b}}_{2})/L$, where $\widehat{\mathbf{b}}_{1} $ and $\widehat{\mathbf{b}}_{2}$ are the reciprocal lattice basis vectors.

Near the critical point and for sufficiently large system sizes, the correlation ratio obeys a finite-size scaling (FSS) form \cite{Fisher71,dosSantos81a,Barber83}, which, by virtue of being a dimensionless quantity, may be written as
\begin{equation}
 R_\gamma(L,V)=\mathcal{F}[(V-V_c)\,L^{1/\nu}],
  \label{eq:RFSS}
\end{equation}
where $\mathcal{F}$ is an unknown function of its argument and $\nu$ is the correlation length critical exponent.  
Therefore, $\mathcal{F}(0)$ is a constant independent of $L$ at the critical point, so that plots of $R_\gamma(V)$ for different $L$'s should cross at $V_c$, thus providing estimates for the critical point; these, in turn, should converge towards the exact value as $L\to\infty$. 

And, finally, we take $\beta = L$ to project the ground state, assuming Lorentz invariance at the QCP's \cite{Assaad2013,Otsuka2016,Parisen2015}.

\section{Results}
\label{sec:results}

\subsection{SM-AFM transition ($U>0$)}
\label{subsec:sm-afm}

\begin{figure}[h]
    \centering
    \includegraphics[scale = 0.5]{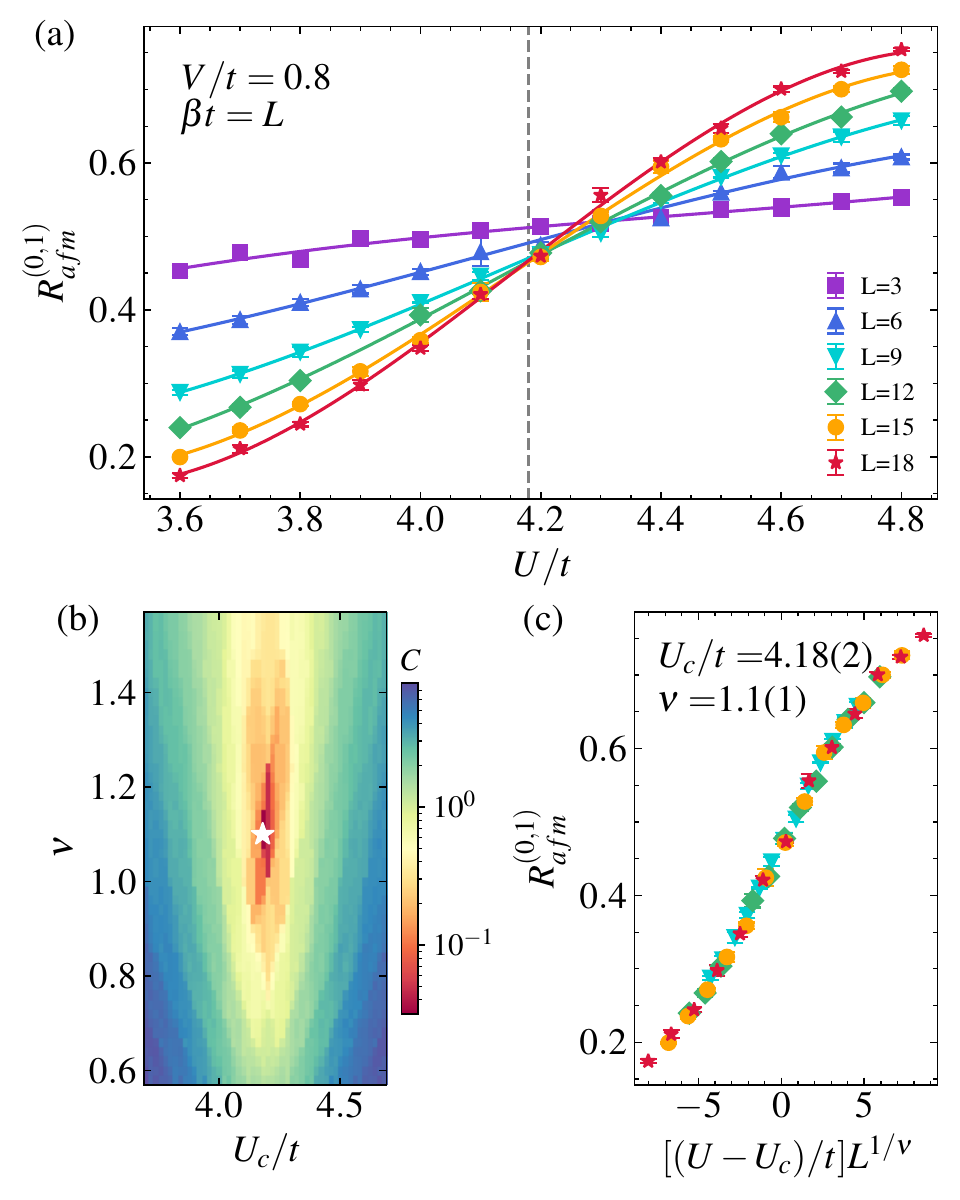}
    \caption{(a) AFM correlation ratio, $R_\text{afm}^{(0,1)}$, at fixed $V$ as a function of $U/t$, for different linear system sizes, $L$. 
    (b) Contour plot of the cost function, $C$, used to fit $R_\text{afm}^{(0,1)}$ to a FSS form, Eq.\,\eqref{eq:RFSS}. 
    The white star locates its  minimum, which in turn determines the best fit to the correlation length exponent, $\nu$, and  $U_c$.
    (c) The resulting FSS data collapse, with fitted $\nu$ and $U_c$; the latter is indicated in (a) by the dashed vertical line.}
    \label{fig:R_afm}
\end{figure}

We start our analysis with the semimetal-antiferromagnetic (SM-AFM) transition. 
On a half-filled square lattice the HM exhibits AFM order in the ground state for any value of $ U > 0 $, due to a nested Fermi surface and a van Hove singularity. 
By contrast, on a half-filled HC lattice, which lacks these properties, the ground state remains semimetallic until a transition to an AFM phase occurs at $(U/t)_c \approx 3.85$~\cite{Assaad2013,Otsuka2016}. 
Since nearest-neighbor repulsion, $V>0$, favors charge ordering, one expects that $(U/t)_c$ increases with increasing $V$. 

In order to investigate AFM order in the presence of a finite $V$, we resort to complex Hubbard-Stratonovich fields to overcome the sign problem  for $|V| \leq |U|/3$, as described in Refs.\,\onlinecite{golor2015,Yao22}.
Results for the AFM correlation ratio with $\delta^{(0,1)}_{\qv}$ are depicted in Fig.\,\ref{fig:R_afm}(a):
for each pair of successive $L$'s, $R_\text{afm}$ cross at slightly different values of $U$. 
We should mention that the situation is similar for $\delta^{(1,1)}_{\qv}$,  with the crossing points spanning across a slightly wider range of values; nonetheless, data for the largest $L$'s provide estimates very close to those obtained from $\delta^{(0,1)}_{\qv}$.  
A precise estimate of $V_c$ can be obtained by invoking the FSS form, Eq.\,\eqref{eq:RFSS}, to plot the data in  Fig.\,\ref{fig:R_afm}(a), but now as a function of $(U-U_c)L^{1/\nu}$, with  
the values of $U_c$ and $\nu$ being determined from the minimization of the cost function, $C$ \cite{Suntajs2020}; see  Fig.\,\ref{fig:R_afm}(b).
The resulting data collapse for $V/t=0.8$ is shown in Fig.\,\ref{fig:R_afm}(c), which yields $U_c/t=4.18 \pm 0.02$ and $\nu=1.1\pm 0.1$.
This estimate of $\nu$ is consistent with the ground state SM-AFM transition for both the HM and the EHM on a half-filled HC lattice belonging to the universality class of the Gross-Neveu model \cite{Gross74,Herbut06,Assaad2013,Parisen2015,Otsuka2016,tang2018}.
This correspondence also leads to a dynamical critical exponent $z=1$ \cite{Hohenberg77}. 
Given that $z=s/\nu$, with $s$ being the gap exponent \cite{Hohenberg77,dosSantos81b,Stinchombe85}, we may expect $s=1$.  
We have carried out similar analyses of the crossing points for other combinations of $U$ and $V$ to obtain the whole SM-AFM critical line  drawn through the (blue) pentagons in  Fig.\,\ref{fig:phase_diagram}.

\subsection{CDW transitions}
\label{subsec:cdw-sm}

\begin{figure}[t]
    \centering
    \includegraphics[scale=0.5]{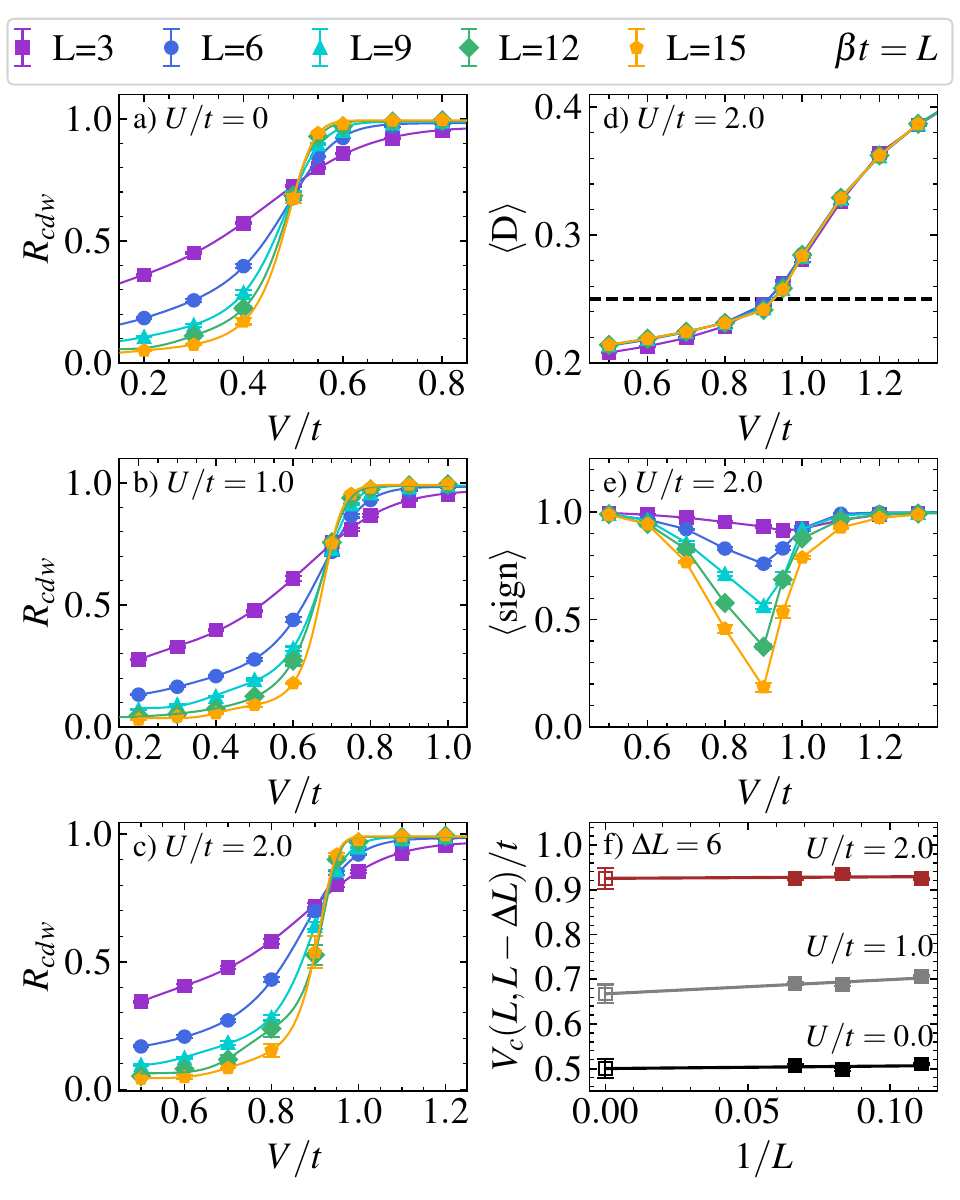}
    \caption{(a)-(c) Charge-density wave correlation ratio for different system linear sizes, $L$, as a function of $V/t$, for fixed $U$.
    (d) Double occupation for $U/t=2$, where the dashed horizontal line indicates the non-interacting value, $\avd=0.25$; see text. 
    (e) Average sign of the product of fermionic determinants for $U/t=2$.
    (f) FSS plots of the crossing points (filled squares) in (a)-(c) using sizes $L$ and $L-\Delta L$, with $\Delta L = 6$, leading to the extrapolated values (empty squares) 
    $V_c/t = 0.50\pm 0.02$, $V_c/t = 0.67\pm 0.02$ and $V_{c}/t = 0.92\pm0.02$, respectively.
    }
    \label{fig:R_cdw}
\end{figure}

In addition to antiferromagnetism, the EHM is expected to display CDW order above some critical curve $V_c(U)>0$, whose shape we will now determine. 
From the outset, we note that the search for CDW order should extend beyond the sign-free region, so that for $|V| > |U|/3$ we use real HS fields and monitor the behavior of $\avs$. 
Figures \ref{fig:R_cdw}\,(a)-\ref{fig:R_cdw}\,(c) show typical plots for $R_\text{cdw}$ as a function of $V$ for increasing values of $U\geq0$; in each panel, data for different system sizes are displayed, whose crossing points provide estimates for the critical values of $V$.  
We note that, although these points lie outside the sign-free region, the calculated averages for $U/t\leq 2$ are not degraded, as indicated by the small error bars.
To reinforce the accuracy of our estimates, in Fig.\,\ref{fig:R_cdw}\,(d) we show the double occupancy, $\avd\equiv\ave{n_{\iv\up}n_{\iv\dn}}$ as a function of $V$, for the same $U$ as in Fig.\,\ref{fig:R_cdw}\,(c); being a local quantity, $\avd$ is less sensitive to both noisy $\avs$ and system size.  
The transition to a CDW state may take place around $\avd=0.25$, the non-interacting value \cite{Nowadnick12,Johnston13,Costa2021}, which in this case occurs at $V_{c}/t = 0.925\pm0.025$ (error estimated by the grid of values of $V$ used).
In addition, Fig.\,\ref{fig:R_cdw}\,(e) shows that while $\avs$ dips, the $\avd$ data in Fig.\,\ref{fig:R_cdw}\,(d) do not degrade, and the dip in $\avs$ sharpens very close to the estimated critical point \cite{Mondaini2022,Mondaini2023,SousaJr2024}. 

A third and thorough determination of the existence of CDW critical points is given by the crossings of its correlation ratio\,\cite{Costa2021}, provided in Figs.\,\ref{fig:R_cdw}\,(a)-\ref{fig:R_cdw}\,(c). By defining $V_c(L,L-\Delta L)$ as the crossing of $R_{\rm cdw}(L)$ and $R_{\rm cdw}(L - \Delta L)$, with $\Delta L = 6$, one may extrapolate such crossings to the thermodynamic limit, as displayed in Fig.\,\ref{fig:R_cdw}\,(f). We have performed similar analyses in the second quadrant, $U<0, V>0$, and we note that $V_c$ for CDW vanishes at $U/t=-3.85$; more on this below.
The points we have obtained for $U/t < 3$ appear as (red) squares in Fig.\,\ref{fig:phase_diagram}.

\begin{figure}[t]
\centering
 \includegraphics[scale=0.5]{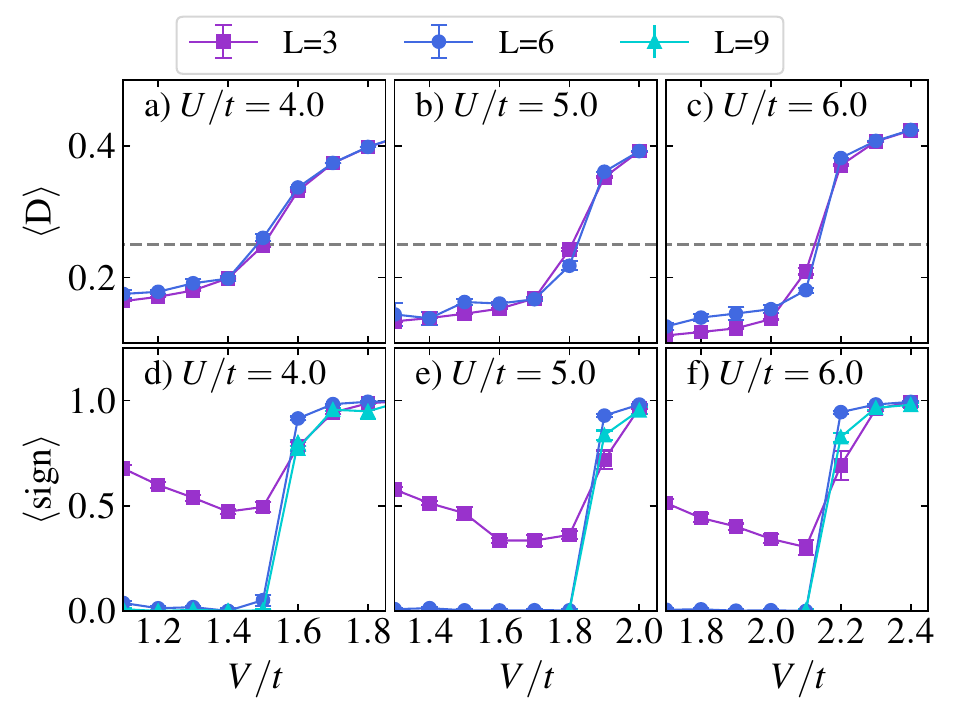}
    \caption{ Double occupancy as a function of $V$ for (a) $U/t = 4$, (b) $U/t=5$ and (c) $U/t=6$. 
    The dashed horizontal lines correspond to $\avd=0.25$. 
    Average determinant sign and a function of $V$ for (d) $U/t = 4$, (e) $U/t=5$ and (f) $U/t=6$. As the system enters the CDW phase, $\avs$ rises sharply.
    }
    \label{fig:DO_cdw}
\end{figure}

For $V>U/3$, $\avs$ decreases as $U$ is increased beyond $U/t=2$; this leads to noisy correlation ratios, to the point of precluding finite-size scaling analyses of critical points. 
However, as discussed above, $\avd$ is less sensitive to both $\avs$ and $L$, so that data for $L\leq 6$ in this region can still provide reliable estimates for $V_c$, especially since $\avd \lesssim 0.25$ in the AFM state. 
Figures \ref{fig:DO_cdw}\,(a)-(c) show the double occupancy for different values of $U$, from which we see that $\avd$ crosses 0.25 at points which hardly depend on $L$.
A totally independent analysis of the transition is provided by the recent observation that a sharp decrease of $\avs$, calculated with real HS fields, can be used to locate quantum critical points \cite{Mondaini2022,Mondaini2023,Wessel2017}; this behavior of $\avs$ at critical points follows from the non-analiticity of the calculated partition function itself\,\cite{Mondaini2022,Mondaini2023}. 
Indeed, Figures \ref{fig:DO_cdw}\,(d)-(f) display $\avs$ for the same values of $U$ as in the upper panels: the points where $\avs$ changes abruptly reproduce with very good accuracy those where $\avd=0.25$.    
The CDW critical points thus obtained appear as (orange) triangles  in Fig.\,\ref{fig:phase_diagram}.

\subsection{SM-SC transition}
\label{subsec:sc-sm}

\begin{figure}[t]
    \centering
    \includegraphics[scale = 0.5]{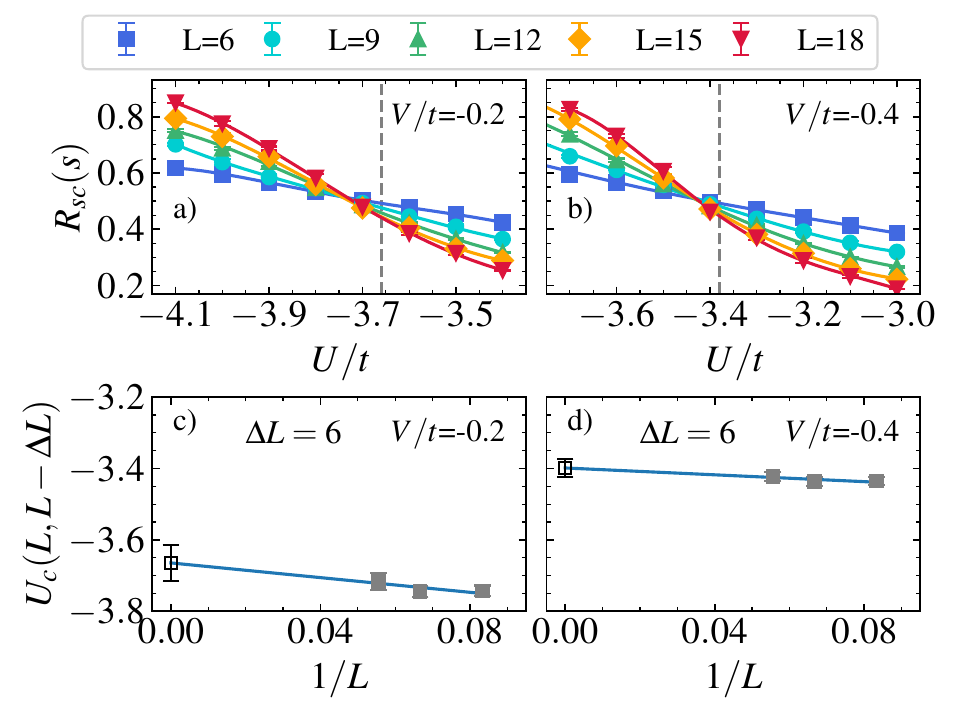}
    \caption{(a) SC correlation ratio as a function of $U$ for $V/t=-0.2$; (b) same as (a), but for $V/t=-0.4$; (c) FSS plot of the crossing points (filled squares) in (a) using sizes $L$ and $L-\Delta$, with $\Delta L=6$, leading to the extrapolated ($L\to\infty$) value $U_{c}/t = -3.67\pm 0.05$ (empty square), indicated by the vertical dashed line in (a);
    (d) same as (c), but for the data in (b), leading to the extrapolated value $U_{c}/t = -3.39\pm 0.02$ respectively.
    }
    \label{fig:Rc_suc}
\end{figure}

Another consequence of the HC lattice being bipartite is that the HM is particle-hole symmetric; therefore, under a spin-down particle-hole transformation with $V=0$, the AFM phase at half filling is mapped onto a half-filled $U<0$ HM in which CDW and superconductivity are degenerate~\cite{Shiba72,Emery76,dosSantos93}. 
Thus, $s$-wave superconductivity occurs at the mirrored region, i.e., for $U/t \lesssim -3.85$ when $V=0$.

For any positive (negative) value of $V$, this degeneracy is broken and the system stabilizes a CDW (superconducting, SC) state.  
More specifically, when $V<0$ we expect an enhancement of pair formation, which, in turn, leads to smaller critical values of $|U|$ needed to induce the SC transition.
In order to investigate the superconducting phase boundary $U_c(V)$, we resort to complex HS fields for $|V| \leq |U| /3$, and consider lattices up to $L=18$. 
Figures \ref{fig:Rc_suc}\,(a) and \ref{fig:Rc_suc}\,(b) show the $s$-wave $R_\text{sc}$ as a function of $U$ for different fixed values of $V$. Similarly to the previous CDW case, we determine the SC critical $U_c$ by examining the crossings of $R_\text{sc}$ for $L$ and $L-\Delta L$, with $\Delta L=6$ to decrease finite-size effects \cite{Costa2021}; the data for $U_c$ thus obtained are then extrapolated to $L\to\infty$, as shown in Figs.\,\ref{fig:Rc_suc}\,(c) and \ref{fig:Rc_suc}\,(d). 
We have used the same procedure for other values of $V$ (not shown) to obtain the critical curve shown as (purple) circles in Fig.\,\ref{fig:phase_diagram}.

We have calculated several other possible pairing structure factors, described by different form factors; see, e.g., Refs.\,\cite{Faye2015,Xu2016,Ying2018}. 
However, we have found no evidence for dominance of any pairing symmetry apart from $s$-wave; see Ref.\,\cite{SousaJr2024} for similar analyses. 
This is in marked contrast with what happens for the square lattice, in which case $d$-wave pairing is stabilized near the non-interacting region \cite{SousaJr2024}.
We again attribute this difference to the `exclusion' of all ordered phases due to the lack of nesting and van-Hove singularity for the half-filled HC lattice.

\begin{figure}[t]
    \centering
    \includegraphics[scale=0.5]{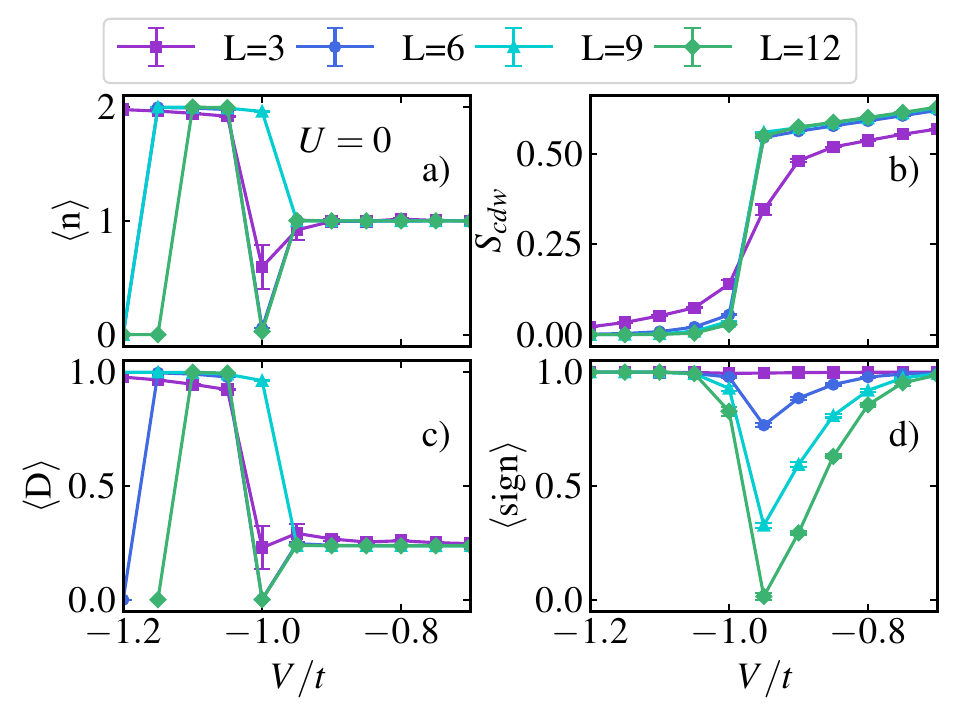}
    \caption{ (a) Average density, (b) CDW structure factor (c) double occupancy, and (d) average sign of fermionic determinants, as functions of $V/t$. 
    All data shown are for $U=0$.}
    \label{fig:ps_u0}
\end{figure}


\subsection{Phase separation}
\label{subsec:ps}

\begin{figure}[t]
    \centering
    \includegraphics[scale = 0.5]{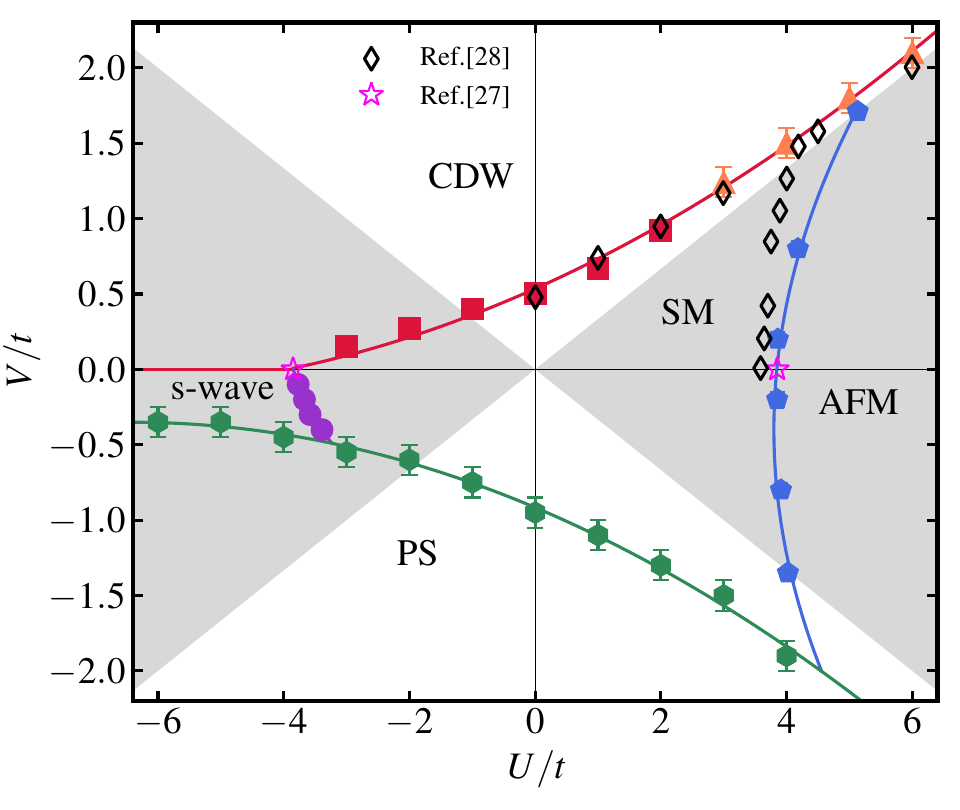}
    \caption{Ground state phase diagram for the Extended Hubbard model on a half-filled honeycomb lattice. 
    Filled symbols are estimates for critical points obtained through our DQMC simulations: (orange) triangles represent results from double occupancy, while (red) squares, (blue) pentagons, and (purple) circles are derived from correlation ratio data. 
    The PS transition points, shown in (green) hexagons, are determined from density, charge structure factor, double occupancy, and $\avs$.
    Empty symbols are critical point estimates from DCA (diamonds, Ref.\,\cite{Wu2014}), and from QMC (stars, Ref.\,\cite{Otsuka2016}). 
    The $|V|\leq |U|/3$ region (shaded) yields sign-free simulations. Lines through data points are guides to the eye.
    }
    \label{fig:phase_diagram}
\end{figure}

For sufficiently strong nearest neighbor attraction, $V<0$, and over a wide range of values of $U$, fermions tend to form clusters of doubly occupied nearest neighbor sites. 
This has a high degeneracy, and is known as a phase separated (PS) phase; it also occurs on the square lattice \cite{SousaJr2024}.
As discussed in Refs.\,\cite{Batrouni2019,SousaJr2024}, one of the features of the PS state is the divergence of the compressibility, due to strong fluctuations in the electronic density, as displayed in Fig.\,\ref{fig:ps_u0}\,(a); it is also manifest in the double occupancy, shown in Fig.\,\ref{fig:ps_u0}\,(c). 
Since the PS ground state is formed by linear combinations of states with $n=0$ and $n=2$, the staggered CDW structure factor drops to zero within the PS region, as shown in Fig.\ref{fig:ps_u0}\,(b).  
Interestingly, $\avs$ also indicates the PS transition through a sharp minimum near $V/t\sim -0.95$, for $U=0$.
By collecting data for  different values of $U$, we map out the whole PS boundary represented by (green) hexagons in Fig.\,\ref{fig:phase_diagram}.

\subsection{Ground state phase diagram}
\label{subsec:gspd}

Our findings for the ground state properties of the EHM on a half-filled HC lattice are summarized in the phase diagram of Fig.\,\ref{fig:phase_diagram}.
We recall that the HM (i.e., $V=0$) on the same lattice displays  three phases, tunable through the strength and sign of the on-site interaction: $s$-wave superconducting coexisting with CDW order, for $U/t\lesssim -3.85$; semi-metallic for $-3.85 \lesssim U_c \lesssim 3.85$; and antiferromagnetic for $U_c \gtrsim 3.85$.
That is, the absence of nesting and a van Hove singularity at half filling prevents the formation of CDW+SC and AFM states; for comparison, we note that for the square lattice, any $U>0$ gives rise to an AFM state, while any $U<0$ gives rise to the degenerate SC and CDW states \cite{SousaJr2024}.
This also occurs when a nearest-neighbor repulsion is added: a purely CDW state is only formed for any $V>0$ above the SC state; for $U\gtrsim -3.85$ the CDW state is only formed above a critical curve $V_c(U)$. 
We also note that increasing $V$ requires larger values of $U$ to stabilize the AFM state; at the same time, for sufficiently large $U$, a direct transition between AFM and CDW takes place, without the need to go through an intermediate SM phase; through the current accuracy of our data, we may expect this to occur for $U/t\gtrsim 6$.
For the square lattice, when $U<0$ a CDW state is formed for any $V>0$, and above some critical curve.
Interestingly, for both lattices the critical curve for $U/t\gg1$ is very close to the strong coupling prediction for the AFM-CDW transition, $V_c=U/z$, where $z$ is the coordination number, which, in addition, corresponds to the boundary of the sign-free region for both lattices, even in the small $U$ region.

The `exclusion' of all ordered phases near the non-interacting region also occurs in the $V<0$ half-plane, being bounded below by a PS region.
In the $U<0$ region, the $s$-wave superconducting phase extends to $V<0$, with $|U_c(V)/t|$ decreasing slightly from $-3.85$; see the (purple) circles in Fig.\,\ref{fig:phase_diagram}. 
However, no change in the symmetry of the paired state was observed: superconductivity is simply suppressed as $U$ is increased, again in marked contrast with the square lattice \cite{SousaJr2024}.
Similarly to what happens in the $V>0$ half-plane, the AFM and PS boundaries seem to merge near $(U,V)\approx(4.2,-2.0)$.

As mentioned before, all previous studies concentrated on the first quadrant \cite{Wu2014,Faye2015,Schuler2018}.  
Results from Dynamical Cluster Approximation \cite{Wu2014} for the SM-CDW and SM-AFM transitions are included in Fig.\,\ref{fig:phase_diagram} for comparison; those from Ref.\,\cite{Faye2015} are very close to the former. 
The agreement with our estimates is very good, although our SM-AFM line is slightly displaced towards larger values of $U$.

\section{Conclusions}
\label{sec:conclusions}

We have studied the extended Hubbard model on a half-filled honeycomb lattice through determinant quantum Monte Carlo simulations to obtain several structure factors characterizing the possible ordered phases.
The `minus-sign problem' in the region $|V|\leq |U|/3$ was avoided with the use of complex Hubbard-Stratonovich auxiliary fields; when necessary, in other regions we have additionally used the double occupancy, and the actual average sign of the product of fermionic determinants. 
Employing careful extrapolations towards zero temperature and thermodynamic limit (with the aid of finite-size scaling) we have proposed a ground state phase diagram covering the whole $(U,V)$ plane, including the hitherto unexplored domains of attractive interactions, $U<0$ and/or $V<0$.

We have established that the lack of nesting and of a van Hove singularity at half-filling produces an `exclusion' region around the non-interacting limit, $U=V=0$, in which no ordered phase is stabilized; this exclusion region corresponds to a semi-metallic (SM) phase, due to the presence of Dirac cones.
Nonetheless, ordered phases may be stabilized beyond critical boundaries.
For instance, in the $V>0$ region we have obtained the critical line running across $U<0$ and $U>0$, above which a charge density wave state may be found. 

Interestingly, while for the square lattice one finds two distinct superconducting regions in the $V<0$ half-plane, namely $s$- and $d$-wave pairing states, here only $s$-wave can be stabilized, and restricted to the $U,V<0$ quadrant; we have examined the behavior of other pairing states, but found that they were suppressed as $T\to 0$ for any set of parameters. 
For sufficiently strong $V<0$ the system exhibits phase separation.

In the $U>0$ half-plane, one finds an antiferromagnetic (AFM) critical line, which asymptotically ($V\approx U/3$ with $U/t\gtrsim 6)$ meets the charge density wave (CDW) line when $V>0$; in this $U,V>0$ region, our phase diagram is similar to the one for the Hubbard-Holstein model on a honeycomb lattice \cite{Costa2021}, thus illustrating a qualitative similarity between the effects of extended electronic interaction and electron-phonon coupling on the properties of honeycomb-based compounds. 
Still in this $U>0$ half-plane, but with $V<0$, the AFM boundary meets the PS phase. By extending the analysis to both repulsive and attractive interaction cases, our results help bridge a gap in the literature regarding the phase diagram of the Hubbard model on the honeycomb lattice.

\section*{ACKNOWLEDGMENTS}
The authors are grateful to the Brazilian Agencies Conselho Nacional de Desenvolvimento Cient\'\i fico e Tecnol\'ogico (CNPq), Coordena\c c\~ao de Aperfei\c coamento de Pessoal de Ensino Superior (CAPES),  and Instituto Nacional de Ci\^encia e Tecnologia de Informa\c c\~ao Qu\^antica (INCT-IQ) for funding this project.
W.K. acknowledges support from FAPERJ Grant No. E-26/203.726/2024 - SEI-260003/011422/2024.
N.C.C.~also acknowledges support from FAPERJ Grant No.~E-26/200.258/2023 - SEI-260003/000623/2023, and CNPq Grant No.\ 313065/2021-7.
R.R.d.S.~acknowledges support from FAPERJ Grant No.\ E-26/210.974/2024, and CNPq Grant No.\ 314611/2023-1.


\bibliography{ref}

\end{document}